# Superconducting properties and electronic structure of CuAl$_2$-Type transition-metal zirconide Fe$_{1-x}$Ni$_x$Zr$_2$

Ryunosuke Shimada[1,2], Yuto Watanabe[1], Lorenzo Tortora[2], Giovanni Tomassucci[2], Muammer Yasin Hacisalihoglu[2,3], Hiroto Arima[1,4], Aichi Yamashita[1], Akira Miura[5], Chikako Moriyoshi[6], Naurang L. Saini[2], Yoshikazu Mizuguchi[1]*

1. Department of Physics, Tokyo Metropolitan University, 1-1, Minami-osawa, Hachioji 192-0397, Japan.
2. Dipartimento di Fisica, Universita´ di Roma "La Sapienza" – P. le Aldo Moro 2, 00185 Roma, Italy.
3. Department of Physics, Recep Tayyip Erdogan University, 53100 Rize, Turkey.
4. National Institute of Advanced Industrial Science and Technology, 1-1-1 Umezono, Tsukuba, Ibaraki 305-8563, Japan
5. Faculty of Engineering, Hokkaido University, Sapporo, Hokkaido 060-8628, Japan.
6. Graduate School of Advanced Science and Engineering, Hiroshima University, Higashihiroshima, Hiroshima 739-8526, Japan

(*mizugu@tmu.ac.jp)

CuAl$_2$-type transition-metal (*Tr*) zirconides are superconductor family, and the *Tr*-site element substitution largely modifies its transition temperature ($T_c$). Here, we synthesized polycrystalline samples of Fe$_{1-x}$Ni$_x$Zr$_2$ by arc melting. From magnetic susceptibility measurements, bulk superconductivity was observed for $0.4 \leq x \leq 0.8$, and the highest $T_c$ of 2.8 K was observed for $x = 0.6$. Specific heat measurements were also performed, bulk superconductivity was observed for $0.4 \leq x \leq 0.8$, and the highest $T_c$ of 2.6 K was observed for $x = 0.6$. The obtained superconductivity phase diagram exhibits dome-shaped trend, which is similar to unconventional superconductors, where magnetic fluctuations are essential for superconductivity. In addition, from the $c/a$ lattice constant ratio analysis, we show the possible relationship between the suppression of bulk superconductivity in the Ni-rich compositions and a collapsed tetragonal transition.

# 1. Introduction

Transition-metal zirconides with a tetragonal CuAl$_2$-type crystal structure (*I4/mcm*, No. 140) are known as a superconductor family with a variable transition temperature ($T_c$) depending on the transition-metal (*Tr*) element [1–9]. The highest $T_c$ is 11.3 K in RhZr$_2$ among the *Tr*Zr$_2$ family, and the superconducting mechanism is basically understood by phonon-mediated conventional mechanism [8,10]. However, importance of antiferromagnetic spin fluctuation to superconductivity was proposed for CoZr$_2$ [6]. In addition to relatively high $T_c$, *Tr*Zr$_2$ compounds have been drawing much attention on anomalous thermal expansion [9,11–14] and high-entropy material design [13,15,16], in which selection of the *Tr*-site element is essential for physical properties. Therefore, to enrich the knowledge on superconducting properties of *Tr*Zr$_2$, further investigation for various combination of elements at the *Tr*Zr$_2$ site is needed.

Among the *Tr*Zr$_2$, the number of studies on superconductivity in FeZr$_2$-related samples is limited. In Ref. 3, Fe$_y$Ni$_{1-y}$Zr$_2$ amorphous metallic-glass samples were investigated for $y \leq$ 0.6, and the highest $T_c$ of ~2.6 K was found for $y$ = 0.1. Furthermore, magnetism and superconducting properties of Fe-Zr binary phases were systematically studied [17,18]. When $z$ in the Fe$_{1-z}$Zr$_z$ metallic glass is lower than 0.6, ferromagnetic ordering is observed, and superconductivity is observed for $z$ > 0.7. In between the ferromagnetic and superconducting phases, paramagnetic phases are present. Because the Zr concentration in the target phase *Tr*Zr$_2$ is 66.7%, FeZr$_2$ is expected to locate in the vicinity of ferromagnetic and superconducting states in the Fe-Zr binary phase diagram. As well known, superconductivity emerging in the vicinity of magnetic ordering possesses unconventional mechanism [19]. As mentioned above, CoZr$_2$ may possess collaboration between superconductivity and antiferromagnetic spin fluctuation. Therefore, FeZr$_2$-based superconductors may be potential unconventional superconductors. The previous studies on FeZr$_2$-based materials were performed on glassy samples prepared by a melt spinning method. In this study, we aimed to investigate physical properties of homogeneous samples of $Fe_{1-x}Ni_xZr_2$ by preparing samples using arc melting. A dome-shaped superconductivity phase diagram with the highest $T_c$ of 2.8 K was obtained for polycrystalline samples of $Fe_{1-x}Ni_xZr_2$.

## 2. Experimental details

Polycrystalline samples of $Fe_{1-x}Ni_xZr_2$ ($x$ = 0, 0.1, 0.2, 0.3, 0.4, 0.5, 0.6, 0.7, 0.8, 0.9, 1.0) were synthesized by arc melting in an Ar atmosphere. Powders of pure transition metals (*Tr*) of Fe (99.9%, Kojundo Kagaku) and Ni (99.9%, Kojundo Kagaku) with a nominal composition were mixed and pelletized. The *Tr* pellet and the plates of pure Zr (99.2%, Nilaco) were used as starting materials. The samples were melted three times and turned over each time for homogenization.

The crystal structure and the purity of the obtained samples were investigated by synchrotron X-ray diffraction (SXRD) at $T$ = 300 K at BL13XU, SPring-8 (proposal No.: 2023A1042). The SXRD data were collected using MYTHEN system [20] using X-ray with a wave length of $\lambda = 0.354032$ Å. Lattice constants were determined by Rietveld refinement using RIETAN-FP [21], and the schematic image of the crystal structure was drawn using VESTA [22]. The actual compositions of the samples were investigated using energy-dispersive X-ray spectrometry (EDX, Swift-ED, Oxford) on a scanning electron microscope (SEM, TM3030Plus, Hitachi Hightech). We measured randomly-selected ten points on the samples surface, and the actual Ni concentration relative to Fe was evaluated by the mean value with standard errors.

XPS measurements were performed on polycrystalline samples of $Fe_{1-x}Ni_xZr_2$ ($x$ = 0, 0.1, 0.2, 0.3, 0.4, 0.5, 0.6, 0.7, 0.8, 0.9, 1.0) at room temperature using the in-house UHV system at the Sapienza University of Rome, equipped with a double-anode XR50 X-ray source and an AR125 Omicron electron analyzer. The base pressure during the measurements was $\sim 3 \times 10^{-9}$ mbar. The Al *Kα* emission line ($hv$ = 1486.6 eV) was used to measure core-level XPS. The X-ray incident and photoelectron emission angles were ~45 deg. for the measurements. The sample surfaces were repeatedly scraped *in situ* by a diamond file to obtain clean surfaces.

The temperature dependences of magnetic susceptibility ($4\pi\chi$) were measured using a superconducting quantum interference device (SQUID) magnetometer (MPMS3, Quantum Design) after both zero-field cooling (ZFC) and field cooling (FC) with an applied field of 1 mT. The temperature dependence of specific heat (*C*) measurements under magnetic fields were performed using a physical property measurement system (PPMS Dynacool, Quantum Design) equipped with a 9 T superconducting magnet. The *C* measurement was carried out by means of a thermal relaxation method. The sample was mounted on a stage with N-grease for good thermal connection.

## 3. Results and discussion

Figure 1(a) shows the SXRD patterns for $x$ = 0–1.0. These compounds have a tetragonal CuAl$_2$-type structure (*I*4/*mcm*, No. 140), and the main peaks could be indexed with the tetragonal structural model. The peaks systematically shift by Ni substitution. Figure 1(c) shows the schematic images of crystal structure of Fe$_{1-x}$Ni$_x$Zr$_2$. Small impurity peaks of the orthorhombic FeZr$_3$ (or *Tr*Zr$_3$) phase are seen as shown in Fig. 1(b); for $x$ = 0.6, the FeZr$_3$ impurity ratio obtained from the two-phase analysis was 11.9 wt.%. Rietveld refinement results for other compositions are shown in Fig. S1. We estimated lattice constants *a* and *c* by the two-phase Rietveld refinements using the SXRD patterns at 300 K, and the obtained *x* dependence of lattice constants are plotted in Fig. 2(a). The Ni concentrations are obtained using EDX. The obtained actual compositions at the *Tr* site which was measure using EDX are comparable to the nominal values as shown in Fig. 3. The *x* dependences of the lattice constants are consistent with the shift of the corresponding SXRD peak positions; hence, the influence of the impurity phases is almost negligible in the evaluation of the changes in lattice constants.

Figure 4(a) shows the XPS spectra for Fe-2p and Ni-2p core level for all the examined samples. We fit the peaks to estimate the compositional ratio of Fe and Ni. The estimated Ni concentration is plotted as a function of nominal $x$ in Fig. 4(b). As indicated by the solid line, which is ideal value from the nominal compositions, the trend of the estimated Ni concentration is consistent with the nominal values, which suggests successful Ni/Fe solution in the examined samples.

Figure 5 shows temperature dependences of magnetic susceptibility of $Fe_{1-x}Ni_xZr_2$ measured in ZFC process. We observed a superconducting transition for $0.4 \leq x \leq 1.0$. There is no multiple-step superconducting transition in the temperature dependence for these compositions between 1.8 and 10 K, which is another proof of homogeneous (systematic) Ni substitution in $Fe_{1-x}Ni_xZr_2$. The large diamagnetic signals observed for $0.4 \leq x \leq 0.8$ suggests the emergence of bulk superconductivity. In contrast, the signals for $x > 0.8$ are clearly small as a bulk superconductor, which indicates that the observed diamagnetic signals are caused by filamentary (trace) superconductivity states in those samples. Above 1.8 K, no superconductivity was observed for $0 \leq x \leq 0.3$. Bulk superconductivity was first observed at $x = 0.4$, and $T_c$ increased until $x = 0.6$. The highest $T_c$ of 2.8 K was observed for $x = 0.6$. However, for $x > 0.6$, the $T_c$ tends to decrease with increasing $x$. Figure 6 shows the $x$ dependence of $T_c$. A dome-shaped superconductivity phase diagram was observed. As mentioned above, the samples with $x > 0.8$ exhibit filamentary superconductivity. In Fig. 6, we indicated the boundary between bulk superconductivity (Bulk SC) and filamentary superconductivity (Filamentary SC).

To further confirm the bulk nature of the observed superconducting transitions for the samples of $Fe_{1-x}Ni_xZr_2$ ($x = 0, 0.4, 0.5, 0.6, 0.7, 0.8, 0.9, 1$), $C$ measurements were performed. Figures 7(a)–(g) show the squared-temperature ($T^2$) dependences of the total $C(T)/T$ under 0 and 9 T. A clear jump was observed for $x = 0.4$–0.8 under 0 T, which indicates bulk nature of their superconducting transitions. No bulk superconducting transition was detected by $C$ for $FeZr_2$ ($x = 0$) and $NiZr_2$ ($x = 1$) under 0 T above 1.8 K. The $C(T)/T$ data under 9 T were fitted to $C(T)/T = \gamma + \beta T^2$, where $\gamma$ and $\beta$ are Sommerfeld coefficient and the coefficient for the phonon contribution to total specific heat, respectively. The estimated $\gamma$ were 18.32(2), 19.68(4), 21.02(4), 19.43(5), 19.45(4), 18.10(7), and 16.03(7) mJ K$^{-2}$ mol$^{-1}$, and $\beta$ were 0.327(1), 0.438(2), 0.466(2), 0.466(3), 0.480(2), 0.478(4) and 0.423(4) mJ K$^{-4}$ mol$^{-1}$ for $x$ = 0, 0.4, 0.5, 0.6, 0.7, 0.8, and 1, respectively. Debye temperature $\Theta_D$ can be calculated using the $\beta$ and the formula $\Theta_D = \left(\frac{12\pi^4 NR}{5\beta}\right)^{1/3}$, where $N = 3$ is the number of atoms per formula unit and $R = 8.31$ J K$^{-1}$ mol$^{-1}$ is an ideal gas constant. The calculated $\Theta_D$ were 261, 237, 232, 232, 230, 230, and 240 K for $x$ = 0, 0.4, 0.5, 0.6, 0.7, 0.8, and 1, respectively. Temperature dependences of the electron contribution of the specific heat $C_{el}(T)$ estimated by subtracting phonon contributions $\beta T^3$ from $C(T)$ are shown in Figs. 7(h)–(j) for $x$ = 0.5, 0.6, and 0.7, respectively. $T_c$ determined from $C_{el}(T)$ under 0 T was 2.5 K, 2.6 K, and 2.5 K for $x$ = 0.5, 0.6, and 0.7, respectively. The normalized jumps of $C_{el}(T)$, $\Delta C_{el}/\gamma T_c$, were estimated as 1.28, 1.30, and 1.35 for $x$ = 0.5, 0.6, and 0.7, respectively. The values of the jump magnitude were similar and slightly lower than 1.43, which is the value expected by the weak-coupling BCS theory [23]. This result suggests that Fe$_{1-x}$Ni$_x$Zr$_2$ ($x$ = 0.5, 0.6, 0.7) are fully-gapped superconductors. When assuming electron-phonon coupling superconductivity, we can calculate an electron-phonon coupling constant $\lambda_{el-ph}$ using the McMillan formula [24]: $\lambda_{el-ph} = \frac{1.04 + \mu^* \ln\left(\frac{\Theta_D}{1.45 T_c}\right)}{(1 - 0.62\mu^*)\ln\left(\frac{\Theta_D}{1.45 T_c}\right) - 1.04}$, where $\mu^* = 0.13$ is a Coulomb coupling constant and the value is used empirically for similar materials containing transition metals. We obtained the values of $\lambda_{el-ph}$ to be 0.57 for $x$ = 0.5, 0.6 and 0.7. An electronic density of states at the Fermi energy $D(E_F)$ is proportional to a term $(1 + \lambda_{el-ph})$ when we consider the electron-phonon coupling. Therefore, $D(E_F)$ with spin degeneracy can be expressed in $D(E_F) = \frac{3\gamma}{\pi^2 k_B^2 (1 + \lambda_{el-ph})}$. The measured $\gamma$ and calculated $\lambda_{el-ph}$ provide $D(E_F)$ = 5.70 states eV$^{-1}$ per formula unit (f.u.), 5.24 states eV$^{-1}$ per f.u., and 5.27 states eV$^{-1}$ per f.u. for $x$ = 0.5, 0.6, and 0.7, respectively. $T_c$ does not increase with an increase in $\Theta_D$ or $\gamma$, unlike the explanation in the BCS theory.

Figure 8(a) shows temperature dependences of total specific heat $C(T)$ at several magnetic fields for $x = 0.6$. Magnetic fields were applied with an increment of 0.2 T up to $\mu_0 H = 1$ T and also measured at $\mu_0 H = 9$ T. Figure 8(b) is the temperature dependences of upper critical field $\mu_0 H_{c2}(T)$ for $x = 0.6$. Then, the upper critical field at 0 K, $\mu_0 H_{c2}(0)$ can be calculated by fitting the data using the Ginzburg-Landau (GL) model [24,25]: $\mu_0 H_{c2}(T) = \mu_0 H_{c2}(0)\left[\frac{1-(T/T_c)^2}{1+(T/T_c)^2}\right]$. We obtained the value of $\mu_0 H_{c2}(0)$ for $x = 0.6$ to be 4.28 T. We found that the value of $\mu_0 H_{c2}(0)$ were lower than that of $\mu_0 H_{c2}(0) = 4.84$ T calculated with $T_c$ of $C(T)$ using the following formula: $\mu_0 H_P = \frac{\Delta(0)}{\sqrt{g}\mu_B} = 1.86 T_c$, where $g = 2$ is a $g$-factor for free electron and $\mu_B \approx 9.27 \times 10^{-24}$ J T$^{-1}$ is a Bohr magneton. The $\Delta(0)$ is a superconducting gap energy at 0 K described as $\Delta(0) = 1.76 k_B T_c$ ($k_B \approx 1.38 \times 10^{-23}$ J K$^{-1}$ is a Boltzmann constant) in the single gap Bardeen−Cooper−Schrieffer (BCS) model [23].

For $x \leq 0.3$, $x \geq 0.9$, bulk nature of superconductivity is suppressed. To explore a possible cause of the suppression of superconductivity, we estimated the $c/a$ ratio of Fe$_{1-x}$Ni$_x$Zr$_2$ using the SXRD data and plotted in Fig. 2(b) as a function of $x$. In Co$_{1-x}$Ni$_x$Zr$_2$, a collapsed tetragonal transition was observed in a Ni-rich region, and bulk nature of superconductivity is suppressed in after the collapsed tetragonal transition [9]. In the case of current system, although $c/a$ linearly decreases with increasing $x$, the slope clearly changes at around $x = 0.1$–0.3 and $x = 0.7$–0.9. For $x = 0$–0.1 and $x = 0.9$–1, another slope can guide the evolution of $c/a$. We consider that the change in the $c/a$ ratio in the Ni-rich region is a kind of transition to the collapsed tetragonal phases as revealed in Co$_{1-x}$Ni$_x$Zr$_2$. Similar collapsed tetragonal transitions were observed in iron-based superconductors CaFe$_2$As$_2$ and KFe$_2$As$_2$ and related layered compound [26–31]. The electronic structure is generally affected by the collapsed tetragonal transition, which affects superconductivity as well [32,33]. On the absence of bulk superconductivity in the Fe-rich region, we have no explanation at present, but we assume that the disappearance of bulk superconductivity would be related to strong spin fluctuations and/or the transition to collapsed tetragonal phase. To clarify that, further structural, electronic, and magnetic properties should be investigated.

4. Summary

We newly synthesized polycrystalline samples of the transition-metal zirconide $Tr$Zr$_2$ superconductor Fe$_{1-x}$Ni$_x$Zr$_2$ by arc melting. From SXRD and XPS measurements, the systematic Fe/Ni solutions were confirmed. From magnetic susceptibility measurements, we observed bulk superconductivity for $0.4 \leq x \leq 0.8$, and the highest $T_c$ of 2.8 K was observed for $x = 0.6$. Specific heat measurements were also performed, bulk superconductivity was observed for $0.4 \leq x \leq 0.8$, and the highest $T_c$ of 2.6 K was observed for $x = 0.6$. Because of the dome-shaped superconductivity phase diagram and the presence of 3d electrons come from magnetic elements of Fe and Ni, possible relationship between superconductivity and magnetic fluctuations is expected, but further experiments are required to address the possible correlation. From $c/a$ ratio analysis, the suppression of bulk superconductivity in the Ni-rich compositions is ascribed as the collapsed tetragonal transition.

**References**

[1] Z. Fisk, R. Viswanathan, G.W. Webb, Solid State Commun. 15 (1974) 1797–1799.
[2] B.T. Matthias, E. Corenzwit, Phys. Rev. 100 (1955) 626–627.
[3] J. Lefebvre, M. Hilke, Z. Altounian, Phys. Rev. B 79 (2009) 184525.
[4] A. Teruya, M. Kakihana, T. Takeuchi, D. Aoki, F. Honda, A. Nakamura, Y. Haga, K. Matsubayashi, Y. Uwatoko, H. Harima, M. Hedo, T. Nakama, Y. Ōnuki, J. Phys. Soc. Jpn. 85 (2016) 034706.
[5] K.J. Syu, S.C. Chen, H.H. Wu, H.H. Sung, W.H. Lee, Physica C 495 (2013) 10–14.
[6] M. Takekuni, H. Sugita, S. Wada, Phys. Rev. B 58 (1998) 11698–11702.
[7] Z. Henkie, W.A. Fertig, Z. Fisk, D.C. Johnston, M.B. Maple, Journal of Low Temperature Physics 48 (1982) 389–403.
[8] K. Yamaya, T. Sambongi, T. Mitsui, J. Phys. Soc. Jpn. 29 (1970) 879.

[9] Y. Watanabe, H. Arima, H. Usui, Y. Mizuguchi, Sci. Rep. 13 (2023) 1008.
[10] C. Tayran, M. Çakmak, Physica B: Condens. Matter 661 (2023) 414904.
[11] Y. Mizuguchi, M.R. Kasem, Y. Ikeda, J. Phys. Soc. Jpn. 91 (2022) 103601.
[12] H. Arima, M.R. Kasem, Y. Mizuguchi, Appl. Phys. Express 16 (2023) 035503.
[13] M.R. Kasem, H. Arima, Y. Ikeda, A. Yamashita, Y. Mizuguchi, J. Phys. Matter. 5 (2022) 045001.
[14] M. Xu, Q. Li, Y. Song, Y. Xu, A. Sanson, N. Shi, N. Wang, Q. Sun, C. Wang, X. Chen, Y. Qiao, F. Long, H. Liu, Q. Zhang, A. Venier, Y. Ren, F. d'Acapito, L. Olivi, D.O. De Souza, X. Xing, J. Chen, Nat. Commun. 14 (2023) 4439.
[15] Y. Mizuguchi, Md.R. Kasem, T.D. Matsuda, Mater. Res. Lett. 9 (2021) 141–147.
[16] M.R. Kasem, A. Yamashita, T. Hatano, K. Sakurai, N. Oono-Hori, Y. Goto, O. Miura, Y. Mizuguchi, Supercond. Sci. Technol. 34 (2021) 125001.
[17] E. Batalla, Z. Altounian, J.O. Strom-Olsen, Phys. Rev. B 31 (1985) 577.
[18] Z. Altounian, J.O. Strom-Olsen, Phys. Rev. B 27 (1983) 4149–4156.
[19] G.R. Stewart, Adv. Phys. 66 (2017) 75–196.
[20] S. Kawaguchi, M. Takemoto, K. Osaka, E. Nishibori, C. Moriyoshi, Y. Kubota, Y. Kuroiwa, K. Sugimoto, Rev. Sci. Instrum. 88 (2017) 085111.
[21] F. Izumi, K. Momma, Solid State Phenom. 130 (2007) 15–20.
[22] K. Momma, F. Izumi, J. Appl. Crystallogr. 41 (2008) 653–658.
[23] L.N. Cooper, J. Bardeen, J.R. Schrieffer, Phys. Rev. 108 (1957) 1175.
[24] W.L. McMillan, Phys. Rev. 167 (1968) 331–344.
[25] A.M. Clogston, Phys. Rev. Lett. 9 (1962) 266–267.
[26] B.S. Chandrasekhar, Appl. Phys. Letters 1 (1962) 7.
[27] A. Kreyssig, M.A. Green, Y. Lee, G.D. Samolyuk, P. Zajdel, J.W. Lynn, S.L. Bud'ko, M.S. Torikachvili, N. Ni, S. Nandi, J.B. Leão, S.J. Poulton, D.N. Argyriou, B.N. Harmon, R.J. McQueeney, P.C. Canfield, A.I. Goldman, Phys. Rev. B 78 (2008) 184517.
[28] A. Van Roekeghem, P. Richard, X. Shi, S. Wu, L. Zeng, B. Saparov, Y. Ohtsubo, T. Qian, A.S. Sefat, S. Biermann, H. Ding, Phys. Rev. B 93 (2016) 245139.
[29] D. Guterding, S. Backes, H.O. Jeschke, R. Valentí, Phys. Rev. B 91 (2015) 140503.
[30] A. Ptok, K.J. Kapcia, M. Sternik, P. Piekarz, J Supercond Nov Magn 33 (2020) 2347–2354.
[31] P.G. Naumov, K. Filsinger, O.I. Barkalov, G.H. Fecher, S.A. Medvedev, C. Felser, Phys. Rev. B 95 (2017) 144106.
[32] R.S. Dhaka, R. Jiang, S. Ran, S.L. Bud'ko, P.C. Canfield, B.N. Harmon, A. Kaminski, M. Tomić, R. Valentí, Y. Lee, Phys. Rev. B 89 (2014) 020511.
[33] A. Ptok, M. Sternik, K.J. Kapcia, P. Piekarz, Phys. Rev. B 99 (2019) 134103.

**Acknowledgements**

The authors thank O. Miura for supports in experiments. This work was partly supported by JSPS-KAKENHI (23KK0088), TMU Research Project for Emergent Future Society, and Tokyo Government-Advanced Research (H31-1). This work has been done under bilateral agreement between Tokyo Metropolitan University and Sapienza University of Rome.

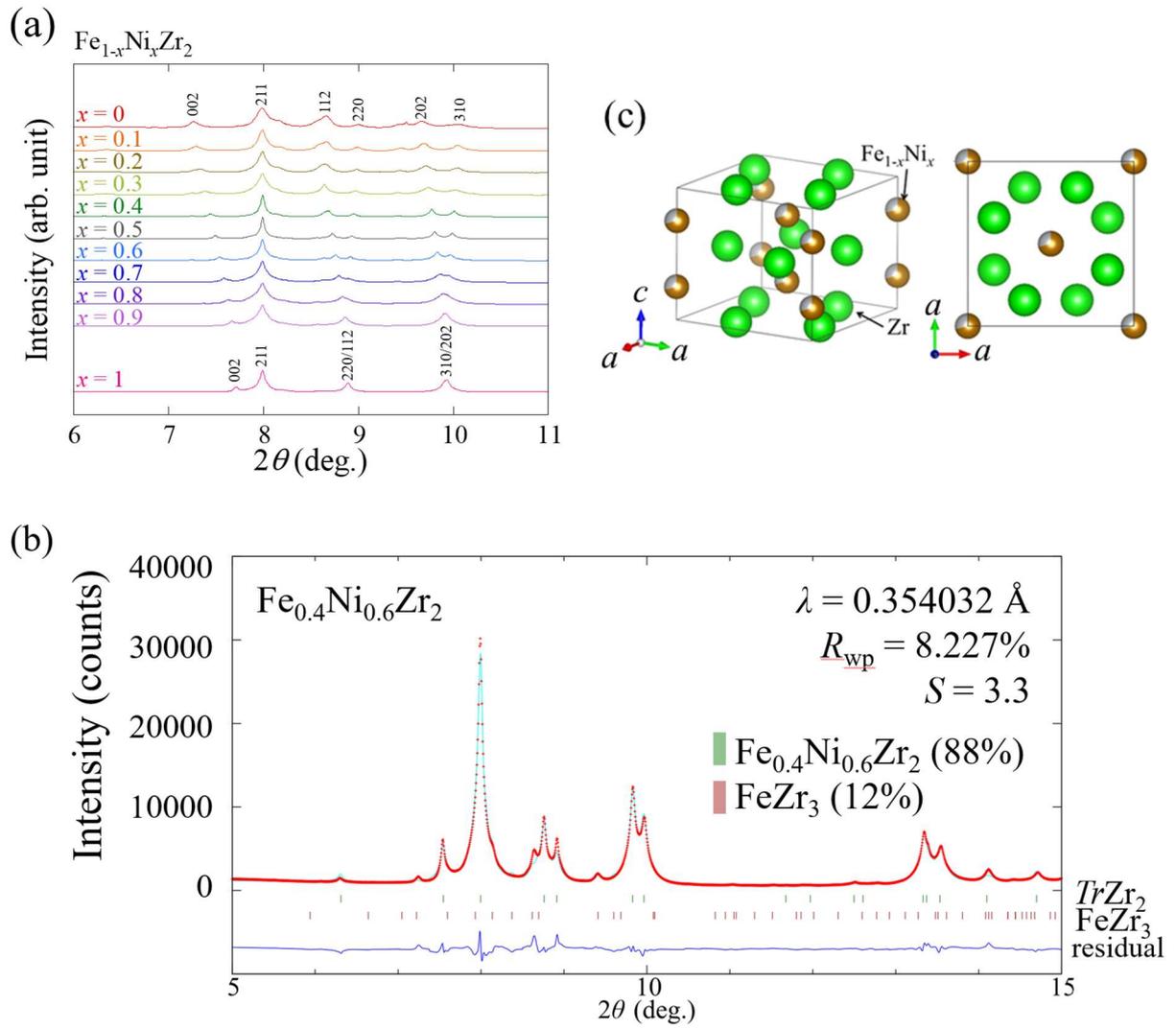

Fig. 1. (a) Powder Synchrotron X-ray diffraction patterns of $Fe_{1-x}Ni_xZr_2$. The numbers are Miller indices. (b) Rietveld refinement result for $x = 0.6$. (c) Schematic images of crystal structure of $Fe_{1-x}Ni_xZr_2$. The impurity contents are shown as mass fraction.

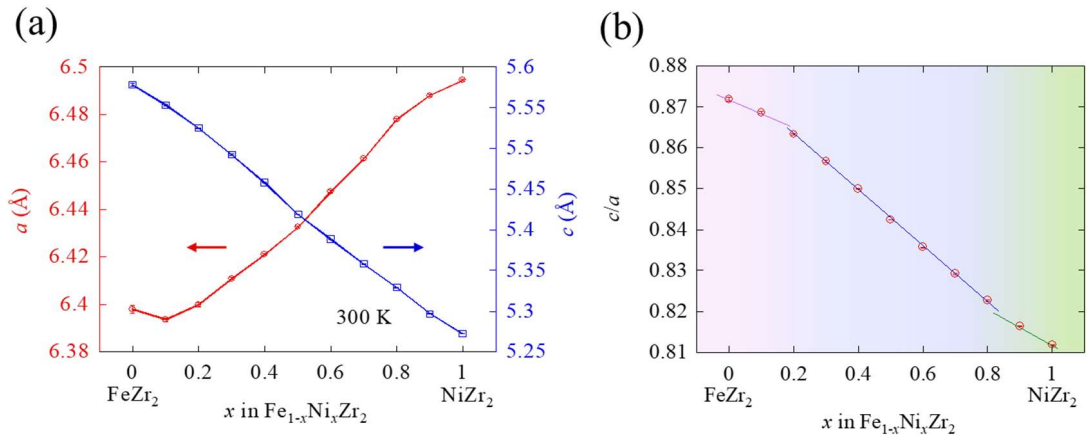

Fig. 2. Ni concentration dependences of lattice constants (a) $a$ and $c$, and (b) $c/a$ ratio of $Fe_{1-x}Ni_xZr_2$. The error bars are standard deviations estimated by the Rietveld refinement.

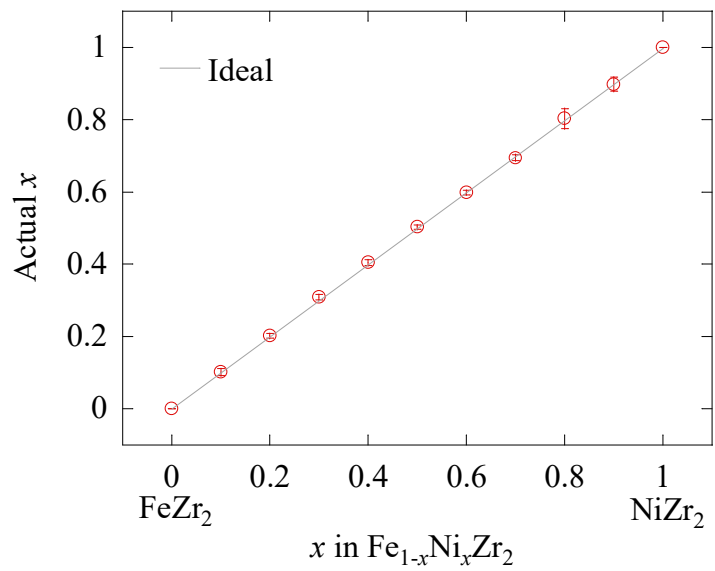

Fig. 3. EDX analysis result: nominal $x$ dependence of actual $x$ in $Fe_{1-x}Ni_xZr_2$. Solid line represents an ideal line when actual $x$ equals nominal $x$.

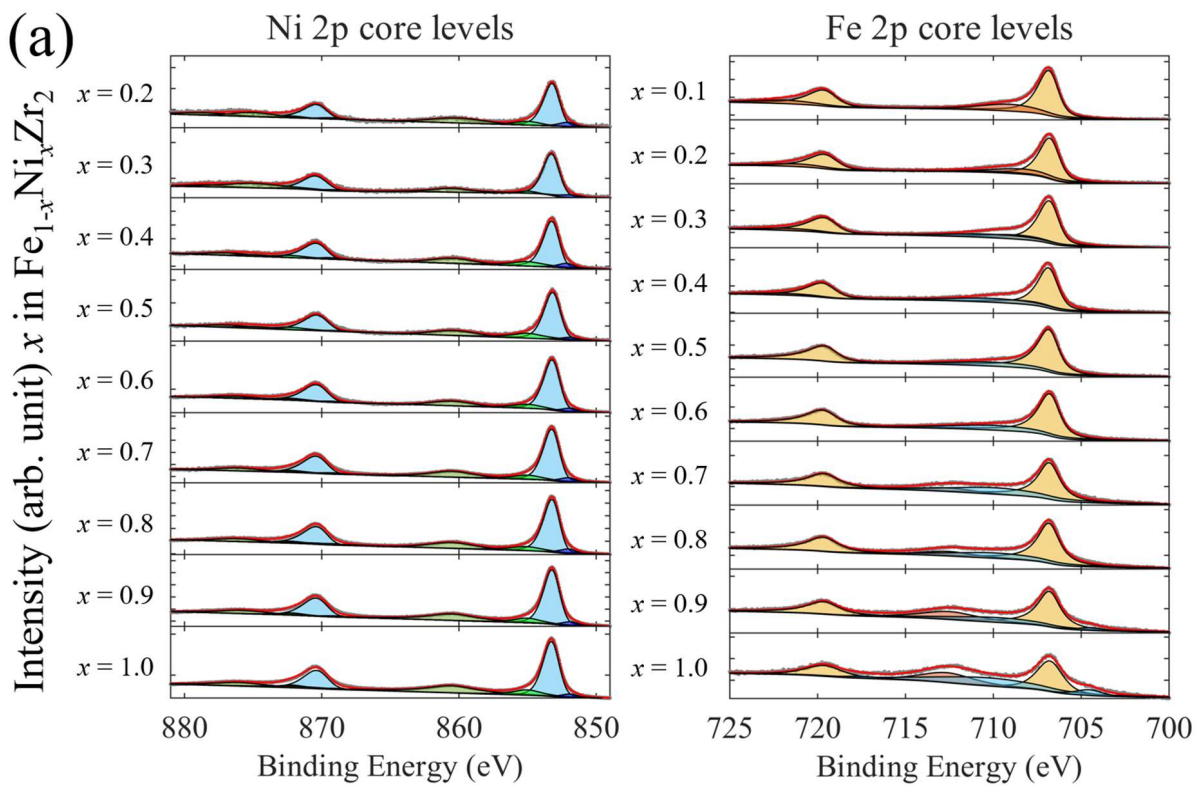

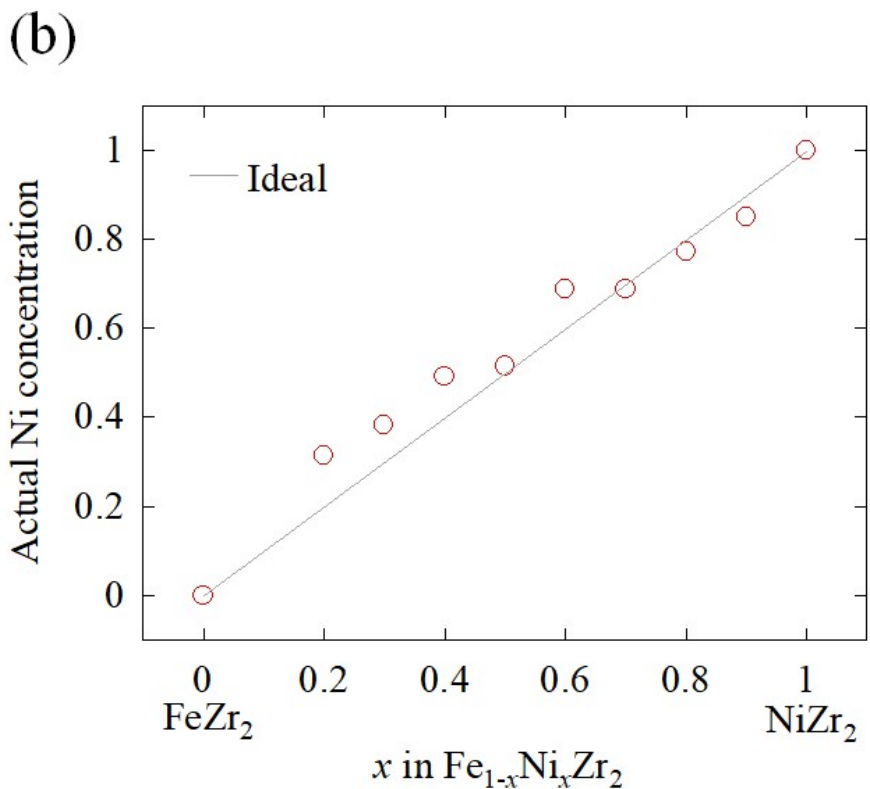

Fig. 4. (a) XPS spectra for Fe-2p and Ni-2p core level. (b) Ni concentration estimated from the XPS analysis. Solid line represents an ideal line when actual $x$ equals nominal $x$.

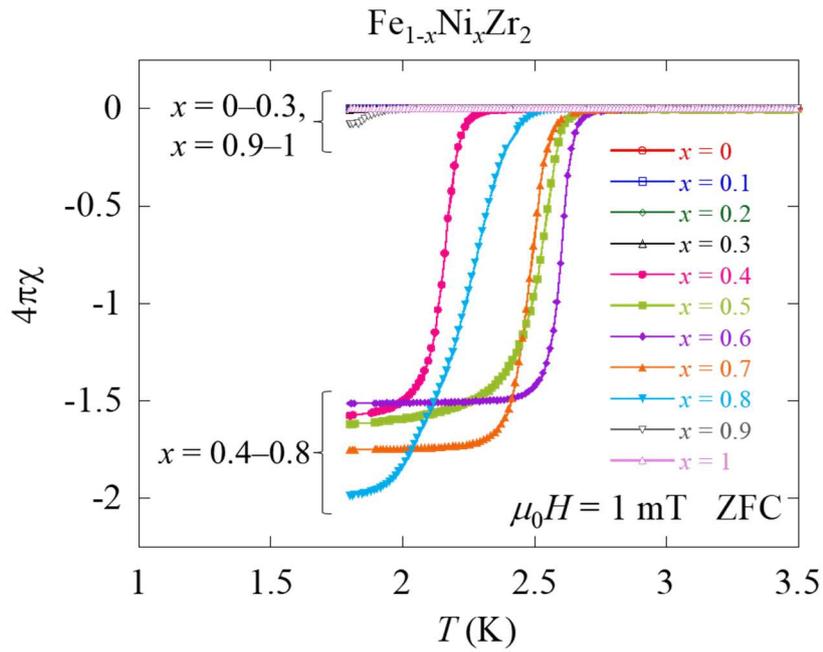

Fig. 5. Temperature dependences of magnetic susceptibility of Fe$_{1-x}$Ni$_x$Zr$_2$.

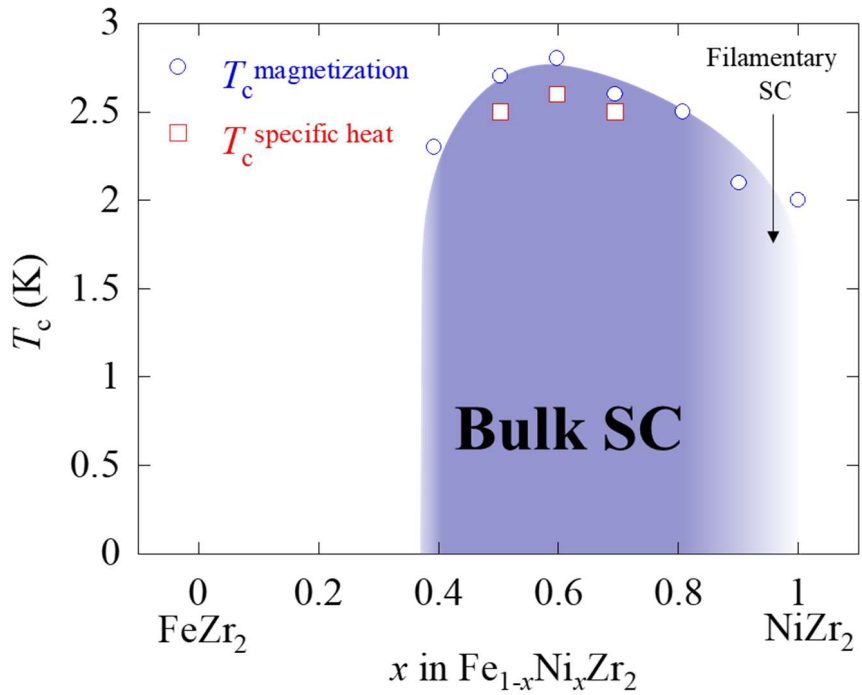

Fig. 6. Superconductivity phase diagram of Fe$_{1-x}$Ni$_x$Zr$_2$.

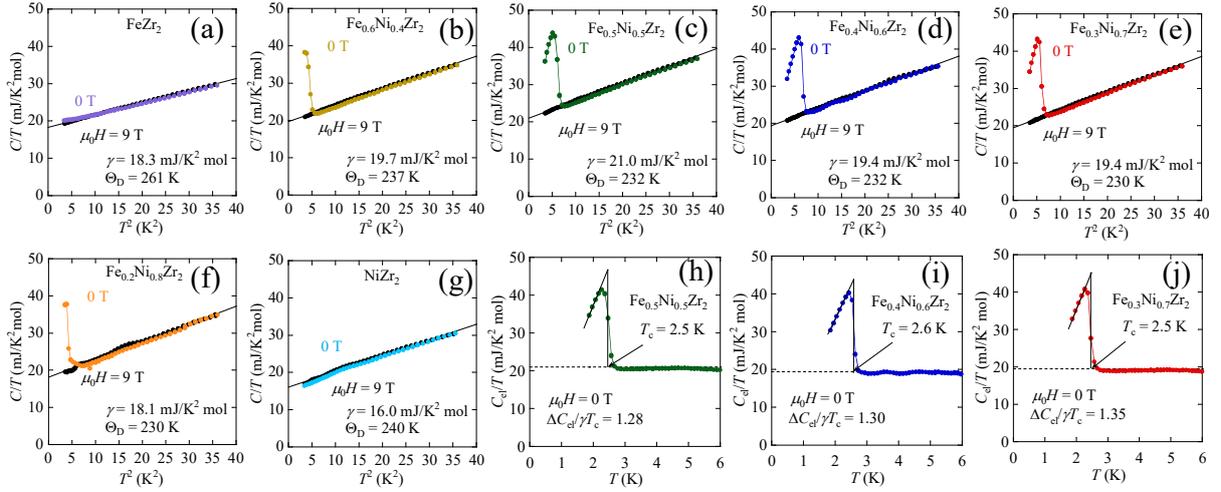

Fig. 7. ((a)–(g)) Squared-temperature ($T^2$) dependences of total specific heat under 0 and 9 T in the form of $C(T)/T$ for Fe$_{1-x}$Ni$_x$Zr$_2$ ($x$ = 0, 0.4, 0.5, 0.6, 0.7, 0.8, 1). The solid lines are fit to $C(T)/T = \gamma + \beta T^2$. ((h)–(j)) Temperature dependences of electronic specific heat under 0 T for $x$ = 0.5, 0.6, 0.7. The solid lines are used to estimate $T_c$ and dashed lines represent $\gamma$ value.

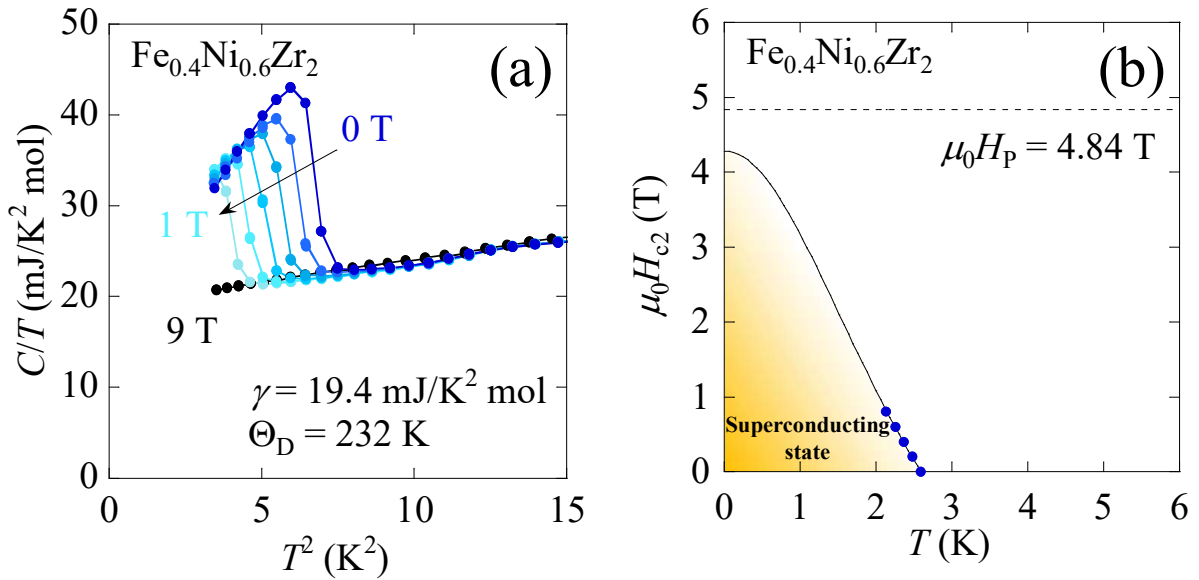

Fig. 8. (a) Temperature dependences of total specific heat under several magnetic fields for Fe$_{0.4}$Ni$_{0.6}$Zr$_2$. (b) Temperature dependence of upper critical field for Fe$_{0.4}$Ni$_{0.6}$Zr$_2$. The solid line is fit to the GL model. The value of $\mu_0 H_P$ was calculated using $\mu_0 H_P = 1.86 T_c$ with $C(T)$ data.

**Supporting information**

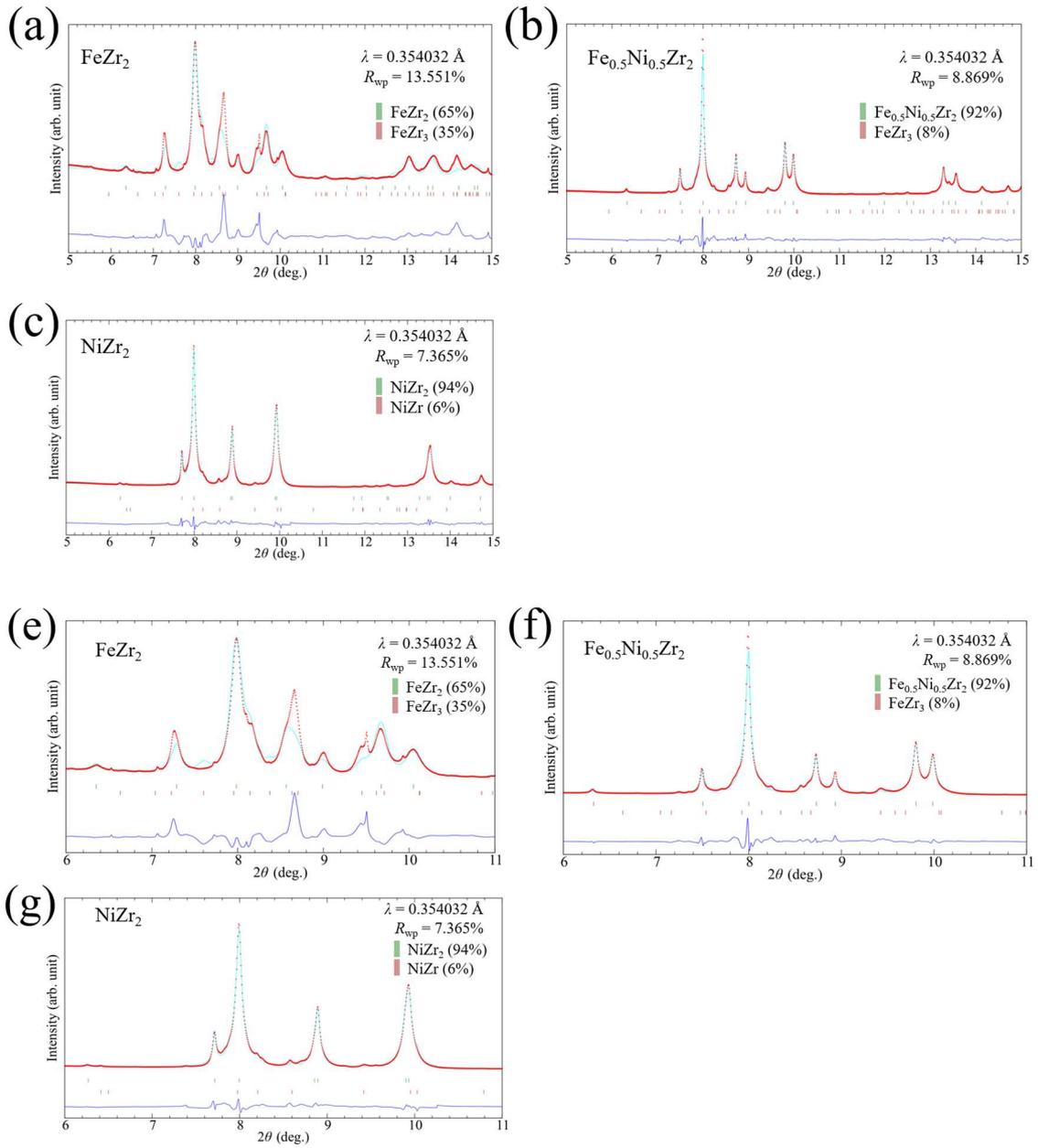

Fig. S1. Rietveld refinement results for $Fe_{1-x}Ni_xZr_2$ ($x$ = 0, 0.5, 1.0).